\documentclass[prd,a4paper,twocolumn]{revtex4}
\usepackage{graphicx}
\usepackage[]{subfigure}

\newcommand{\nc}{\newcommand}

\nc{\be}[1]{\begin{equation}\mbox{$\label{#1}$}}
\nc{\bea}[1]{\begin{eqnarray} \mbox{$\label{#1}$}}
\nc{\Section}[2]{\section{#2}\label{#1}}
\nc{\Bibitem}[1]{\bibitem{#1}}
\nc{\Label}[1]{\label{#1}}

\nc{\eea}{\end{eqnarray}}
\nc{\ee}{\end{equation}}

\nc{\bdm}{\begin{displaymath}}
\nc{\edm}{\end{displaymath}}
\nc{\dpsty}{\displaystyle}
\nc{\bc}{\begin{center}}
\nc{\ec}{\end{center}}
\nc{\ba}{\begin{array}}
\nc{\ea}{\end{array}}
\nc{\bab}{\begin{abstract}}
\nc{\eab}{\end{abstract}}
\nc{\btab}{\begin{tabular}}
\nc{\etab}{\end{tabular}}
\nc{\bit}{\begin{itemize}}
\nc{\eit}{\end{itemize}}
\nc{\ben}{\begin{enumerate}}
\nc{\een}{\end{enumerate}}
\nc{\bfig}{\begin{figure}}
\nc{\efig}{\end{figure}}

\nc{\arreq}{&\!=\!&}
\nc{\arrmi}{&\!-\!&}
\nc{\arrpl}{&\!+\!&}
\nc{\arrap}{&\!\!\!\approx\!\!\!&}
\nc{\non}{\nonumber}
\nc{\align}{\!\!\!\!\!\!\!\!&&}

\def\lsim{\; \raise0.3ex\hbox{$<$\kern-0.75em
      \raise-1.1ex\hbox{$\sim$}}\; }
\def\gsim{\; \raise0.3ex\hbox{$>$\kern-0.75em
      \raise-1.1ex\hbox{$\sim$}}\; }

\nc{\DOT}{\hspace{-0.08in}{\bf .}\hspace{0.1in}}
\nc{\Laada}{\hbox {$\sqcap$ \kern -1em $\sqcup$}}
\nc\loota{{\scriptstyle\sqcap\kern-0.55em\hbox{$\scriptstyle\sqcup$}}}
\nc\Loota{{\sqcap\kern-0.65em\hbox{$\sqcup$}}}
\nc\laada{\Loota}
\nc{\qed}{\hskip 3em \hbox{\BOX} \vskip 2ex}

\nc{\real}{{\rm I \! R}}
\nc{\Z}{{\sf Z \!\!\! Z}}
\nc{\complex}{{\rm C\!\!\! {\sf I}\,\,}}
\def\bigid{\leavevmode\hbox{\small1\kern-3.8pt\normalsize1}}
\def\id{\leavevmode\hbox{\small1\kern-3.3pt\normalsize1}}
\nc{\slask}{\!\!\!/}
\nc{\bis}{{\prime\prime}}
\nc{\pa}{\partial}
\nc{\na}{\nabla}
\nc{\ra}{\rangle}
\nc{\la}{\langle}
\nc{\goto}{\rightarrow}
\nc{\swap}{\leftrightarrow}

\nc{\EE}[1]{ \mbox{$\cdot10^{#1}$} }
\nc{\abs}[1]{\left|#1\right|}
\nc{\at}[2]{\left.#1\right|_{#2}}
\nc{\norm}[1]{\|#1\|}
\nc{\abscut}[2]{\Abs{#1}_{\scriptscriptstyle#2}}
\nc{\vek}[1]{{\rm\bf #1}}
\nc{\integral}[2]{\int\limits_{#1}^{#2}}
\nc{\inv}[1]{\frac{1}{#1}}
\nc{\dd}[2]{{{\partial #1}\over{\partial #2}}}
\nc{\ddd}[2]{{{{\partial}^2 #1}\over{\partial {#2}^2}}}
\nc{\dddd}[3]{{{{\partial}^2 #1}\over
    {\partial #2 \partial #3}}}
\nc{\dder}[2]{{{d #1}\over{d #2}}}
\nc{\ddder}[2]{{{d^2 #1}\over{d {#2}^2}}}
\nc{\dddder}[3]{{d^2 #1}\over
    {d #2 d #3}}
\nc{\dx}[1]{d\,^{#1}x}
\nc{\dy}[1]{d\,^{#1}y}
\nc{\dz}[1]{d\,^{#1}z}
\nc{\dl}[1]{\frac{d\,^{#1}l}{(2\pi)^{#1}}}
\nc{\dk}[1]{\frac{d\,^{#1}k}{(2\pi)^{#1}}}
\nc{\dq}[1]{\frac{d\,^{#1}q}{(2\pi)^{#1}}}

\nc{\bfT}{{\bf T }}

\nc{\cA}{{\cal A}}
\nc{\cB}{{\cal B}}
\nc{\cD}{{\cal D}}
\nc{\cE}{{\cal E}}
\nc{\cG}{{\cal G}}
\nc{\cH}{{\cal H}}
\nc{\cL}{{\cal L}}
\nc{\cO}{{\cal O}}
\nc{\cT}{{\cal T}}
\nc{\cN}{{\cal N}}
\nc{\cR}{{\cal R}}
\nc{\rvac}[1]{|{\cal O}#1\rangle}
\nc{\lvac}[1]{\langle{\cal O}#1|}
\nc{\rvacb}[1]{|{\cal O}_\beta #1\rangle}
\nc{\lvacb}[1]{\langle{\cal O}_\beta #1 |}
\nc{\bb}{\bar{\beta}}
\nc{\bt}{\tilde{\beta}}
\nc{\ctH}{\tilde{\cal H}}
\nc{\chH}{\hat{\cal H}}
\nc{\1}{\aa}
\nc{\2}{\"{a}}
\nc{\3}{\"{o}}
\nc{\4}{\AA}
\nc{\5}{\"{A}}
\nc{\6}{\"{O}}
\nc{\al}{\alpha}
\nc{\g}{\gamma}
\nc{\Del}{\Delta}
\nc{\e}{\textrm{e}}
\nc{\eps}{\epsilon}
\nc{\lam}{\lambda}
\nc{\Om}{\Omega}
\nc{\ve}{\varepsilon}
\nc{\mn}{{\mu\nu}}
\nc{\vp}{\varphi}

\nc{\rf}[1]{(\ref{#1})}
\nc{\nn}{\nonumber \\*}
\nc{\bfB}{\bf{B}}
\nc{\bfv}{\bf{v}}
\nc{\bfx}{\bf{x}}
\nc{\bfy}{\bf{y}}
\nc{\vx}{\vec{x}}
\nc{\vy}{\vec{y}}
\nc{\oB}{\overline{B}}
\nc{\oI}{\overline{I}}
\nc{\oR}{\overline{R}}
\nc{\rar}{\rightarrow}
\nc{\ti}{\times}
\nc{\slsh}{\hskip-5pt/}
\nc{\sm}{Standard~Model~}
\nc{\MP}{M_{\rm Pl}}
\nc{\mpl}{M_{\rm Pl}}
\nc{\tp}{t_{\rm Pl}}

\nc{\pmin}{p_{\rm min}}
\nc{\pmax}{p_{\rm max}}
\nc{\fo}{f_0}
\nc{\foi}{f_{0,i}\,}
\nc{\fop}{f_0^P}
\nc{\fou}{f_0^U}

\nc{\eff}{{\rm eff}}
\nc{\MT}{M_{\rm T}}
\nc{\ML}{M_{\rm L}}
\nc{\kk}{\vek{k}}
\nc{\pp}{{\rm p}}
\nc{\pt}{\partial_t}
\nc{\half}{{1\over 2}}
\nc{\w}{\omega}
\nc{\uhat}{\hat{U}_\w}

\nc{\etal}{\mbox{\it et al.}}
\nc{\ie}{{\it i.e. }}
\nc{\eg}{{\it e.g. }}
\nc{\trh}{T_{\rm RH}}
\nc{\ad}{{a'\over a}}
\nc{\bd}{{b'\over b}}
\nc{\Rd}{{R'\over R}}
\nc{\diag}{{\textrm{diag}}}
\nc{\mato}[1]{\tilde{#1}}
\nc{\sinn}{\textrm{sinn}}
\nc{\sech}{\textrm{sech}}
\nc{\I}{\textrm{I}}
\nc{\II}{\textrm{II}}
\nc{\III}{\textrm{III}}
\nc{\vev}[1]{\langle #1 \rangle}
\nc{\hyp}{\,\; F_{1{\hskip -16pt}2}{\hskip 11pt}}
\nc{\brhom}{\overline{\rho}_M}
\nc{\brho}{\overline{\rho}}
\nc{\rhob}{\overline{\rho}}
\nc{\Pb}{\overline{P}}
\nc{\bH}{\overline{H}}
\nc{\ep}{{1+4\eps}}

\def\smiley{\hbox{\large$\bigcirc$\hspace{-.80em}%
\raise.2ex\hbox{$\cdot\cdot$}\kern-.61em    
\lower.2ex\hbox{\scriptsize$\smile$}}\ }

\def\frowney{\hbox{\large$\bigcirc$\hspace{-.80em}%
\raise.2ex\hbox{$\cdot\cdot$}\kern-.635em
\lower.2ex\hbox{\scriptsize$\frown$}}\ }

\begin{document}

\title{
Anti-Neutrino imprint in Solar Neutrino Flare}
\author{D. Fargion}
\email{daniele.fargion@roma1.infn.it} \affiliation{Physics
Department, Rome Univ.1 and INFN, Ple.A.Moro 2, 00185, Italy}

\date{\today}

\begin{abstract}
Future neutrino detector at Megaton mass might enlarge the
neutrino telescope thresholds revealing cosmic supernova
background and largest solar flares neutrino. Indeed the solar
energetic ($ E_p> 100 MeVs $) flare particles (protons, $\alpha$)
while scattering among themselves  on Solar corona atmosphere must
produce prompt charged pions, whose chain decays are source of  a
solar (electron-muon) neutrino "flare" (at tens or hundreds MeV
energy). These brief (minutes) neutrino "burst" at largest flare
peak may overcome by three to five order of magnitude the steady
atmospheric neutrino noise on the Earth, possibly leading to their
detection above detection thresholds (in a full mixed  three
flavor state). Moreover the birth of $anti-neutrinos$ at a few
tens MeVs is well loudly flaring above a $null$ thermal "hep"
anti-neutrino solar background and also above a tiny supernova
relic and atmospheric noise. The largest prompt solar
anti-neutrino "burst" may be well detected in future Super
Kamikande (Gadolinium implemented) anti-neutrino
$\overline{\nu_e}$ signatures mostly in inverse Beta decay
($\overline{\nu_e} + p \rightarrow n + e^+$) and rarely in higher
energy $muon$ , or even rarest $tau$ neutrino leptons. Our
estimate for the recent and exceptional October - November $2003$
solar flares and January $20th$ 2005 exceptional flare might lead
to a few events above or near unity for existing Super-Kamiokande
and above unity for Megaton detectors. The $\nu$ spectra may
reflect in a subtle way the neutrino flavor oscillations and
mixing in flight.  A comparison of the solar neutrino flare (at
their birth place on Sun and after oscillation on the arrival on
the Earth) with other neutrino foreground is estimated: it offers
an independent track to disentangle the neutrino flavor puzzles
and its most secret mixing angles. The sharpest noise-free
anti-neutrino imprint maybe its first clean voice.
\end{abstract}

\maketitle

\begin{figure}
\includegraphics[width=5cm,height=5cm,angle=0]{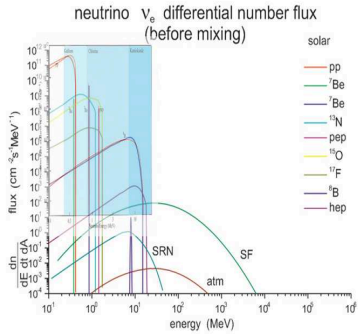}
 \caption{\label{fig:fig1} The multi-component Neutrino flux due to Solar Nuclear Neutrino spectra and lines
  within the atmospheric (atm) Neutrino noises and Supernova Relic
  Spectra (SNR) (or DSNB, Diffuse Supernova Neutrino Background).
   The well known Bachall Solar Neutrino Spectra \cite{ref5}
  overlaps  an  updated and wider neutrino spectra, containing
  the expected  Solar Flare (SF) brightest fluxes \cite{ref0}}.
\end{figure}

\begin{figure}
\includegraphics[width=5cm,height=5cm,angle=0]{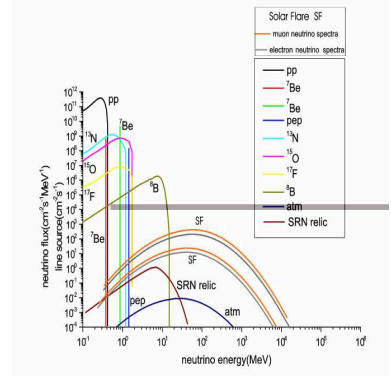}
 \caption{\label{fig:fig1} As above, the ${\nu_e}$  signals of the Solar Flare $SF$ are shown after the flavor mixing
  for two extreme solar neutrino flare luminosity \cite{ref0}: just above or below
  the SK detection threshold.  }
\end{figure}

\begin{figure}
\includegraphics[width=5cm,height=5cm,angle=0]{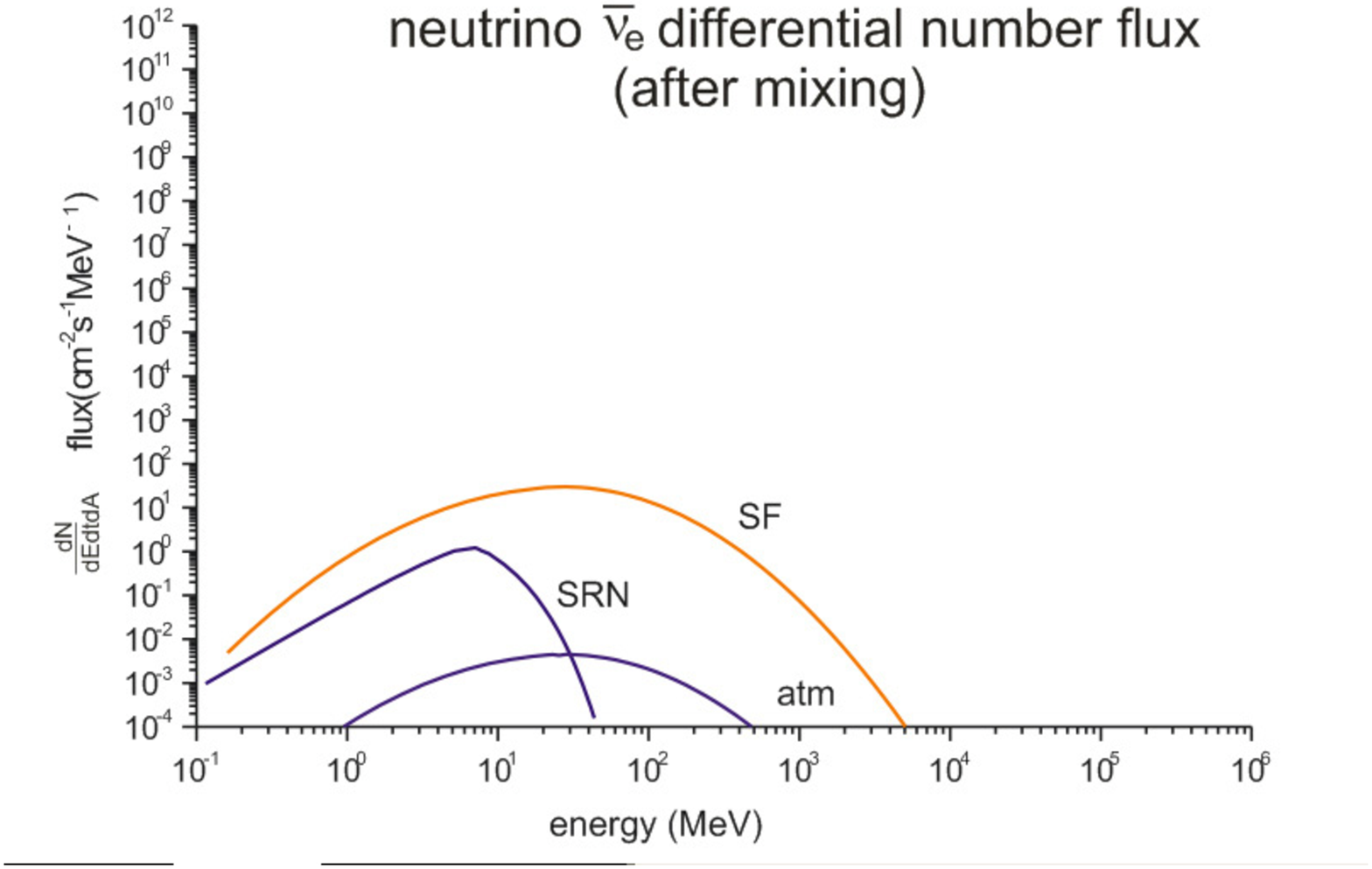}
\caption{\label{fig:fig1}As above  SF $\overline{\nu_e}$
 in noise-free spectra. }
\end{figure}

\begin{figure}
\includegraphics[width=5cm,height=5cm,angle=0]{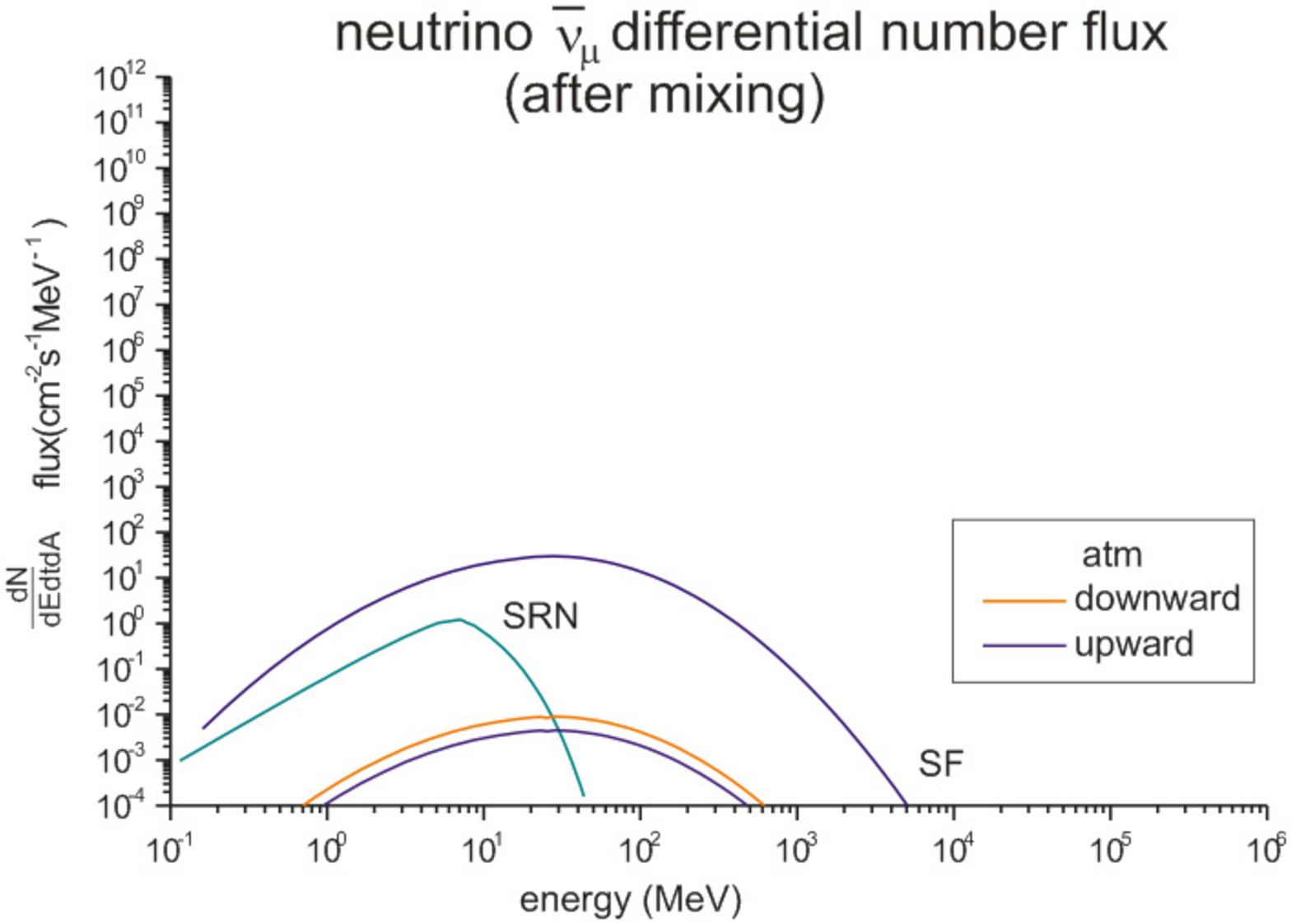}
\caption{\label{fig:fig1}As above for the $\overline{\nu_{\mu}}$
and $\overline{\nu_{\tau}}$ in noise-free background. Up-down
atmospheric $\overline{\nu_{\mu}}$ asymmetry is also shown }
\end{figure}

\section{ Neutrino-Anti-Neutrino Solar Flare}
If during largest solar flare, of a few minutes duration, (1) the
particle flux escaping the corona eruption and hitting later on
the Earth, is 3-4 order of magnitude above the common atmospheric
CR background (while showing similar spectra) and (2) if the flare
particle interactions on the Sun corona is taking place as
efficiently as  in terrestrial atmosphere, than their secondaries
by charged pions and muons decays, are leading to a neutrinos
fluency on Earth comparable to one day terrestrial atmospheric
neutrino activity \cite{ref7}. One therefore may expect a prompt
increase of neutrino signals of the order of one day integral
events made by atmospheric neutrinos. In SK detector the signal is
just on the edge, but as long as the author is aware, it has been
never revealed. Sun density at the flare eruption  might be
diluted and pion production maybe somehow suppressed (by a factor
(0.1-0.05)). This may be the reason for the null detection of huge
event on Oct-Nov. 2003 and Jan. 2005.  Unfortunately the neutrino
signal at hundred MeV energies is rare
  while the one at ten MeV or below is polluted by Solar Hep neutrinos or, in SK,
  by local neutrino reactor noise. Therefore Solar Flare expected signal (at softer spectra)
   maybe dominated by $30$ MeV neutrinos (made by non-relativistic pion decays) that
  might be better revealed by their  anti-neutrino signature: it will be enhanced by
  Gadolinium presence in next SK detectors \cite{Hasan}.   Indeed
  Anti-neutrino electron imprint in Solar Flare may be well
  observable while searching for supernova
  relic anti-neutrino traces in the Universe \cite{Hasan}. The prompt SF signal
  may overcome the atmospheric noise just as the Tau Air-shower
  may overcome atmospheric muons \cite{Fargion 2004}.

There are two different SF (Solar Flare) neutrinos(see
\cite{ref0}): A brief and sharp solar flare, originated within the
$solar$ corona and a diluted and delayed $ terrestrial$ neutrino
flux, produced by flare particles hitting the  Earth's atmosphere.
We consider only the first. The main source of pion production is
$p+p\rightarrow {{\Delta}^{++}}n\rightarrow p{{\pi}^{+}}n$;
$p+p\rightarrow{{{\Delta}^{+}}p}^{\nearrow^{p+p+{{\pi}^{0}}}}_{\searrow_{p+n+{\pi}^{+}}}$
at the center of mass of the resonance ${\Delta}$ (whose mass
value is ${m}_{\Delta}=1232$ MeV). As  a first approximation the
total pion $\pi^+ $ energy is equally distributed, in average, in
all its final remnants: ($\bar{\nu}_{\mu}$, ${e}^{+}$,
${\nu}_{e}$, ${\nu}_{\mu}$):${E}_{{\nu}_{\mu}} \geq
{E}_{{\bar{\nu}_{\mu}}} \simeq {E}_{{\nu}_{e}} \simeq
\frac{1}{4}{E}_{{\pi}^{+}}$.  The flavor oscillation will lead to
a decrease in the muon component and to a hardening  of electron
neutrino component spectra. While at the birth place the neutrino
fluxes by positive charged pions $\pi^+$ are
$\Phi_{\nu_e}$:$\Phi_{\nu_{\mu}}$:$\Phi_{\nu_{\tau}}$ $= 1:1:0$,
after the mixing assuming a  number redistribution we expect
$\Phi_{\nu_e}$:$\Phi_{\nu_{\mu}}$:$\Phi_{\nu_{\tau}}$ $=
(\frac{2}{3}):(\frac{2}{3}):(\frac{2}{3})$. On the other side for
the anti-neutrino fluxes we expect at the birth place:
$\Phi_{\overline{\nu_e}}$:$\Phi_{\overline{\nu_{\mu}}}$:$\Phi_{\overline{\nu_{\tau}}}$
$= 0:1:0$ while at their arrival (within a similar flavor
redistribution)
:$\Phi_{\overline{\nu_e}}$:$\Phi_{\overline{\nu_{\mu}}}$:$\Phi_{\overline{\nu_{\tau}}}$
$= (\frac{1}{3}):(\frac{1}{3}):(\frac{1}{3})$.
We  considered the solar flare neutrino events due to these number
fluxes following known $\nu$-nucleons cross-sections at these
energies \cite{bemporad},\cite{strumia},\cite{Bodek} at
Super-Kamiokande II, finding a detectable signal: $ {N}_{ev}
\simeq 7.5 \cdot \eta \left( \frac{E_{FL}}{10^{31} \, erg} \right)
$ (for more details and explanations see \cite{ref0}). However the
event expectation numbers at SK-II for {\em solar neutrino burst}
assuming a more pessimistic detector thresholds calibrated with
the observed Supernova 1987A event fluxes \cite{ref0} is just at
the detection edge compatible with null SK discover:
$ {N_{ev}}_{\bar{\nu}_{e}} \simeq
0.63{\eta}(\frac{\bar{E}_{\bar{\nu}_{e}}}{35
~MeV})(\frac{E_{FL}}{10^{31}~erg});~\bar{E}_{\bar{\nu}_{e}}\leq
100 ~MeV $;
$ {N_{ev}}_{\bar{\nu}_{e}} \simeq
1.58{\eta}(\frac{E_{FL}}{10^{31}~erg});~
\bar{E}_{\bar{\nu}_{e}}\geq100-1000 ~MeV $;
$ {{N}_{ev}}_{\bar{\nu}_{\mu}} \simeq
3.58{\eta}(\frac{{E}_{FL}}{{10}^{31}~erg});~
\bar{E}_{\bar{\nu}_{\mu}}\geq 200-1000 ~MeV $;
where the efficiency factor ${\eta}\leq1$.
   The background due to
energetic atmospheric neutrinos at the Japanese detector is nearly
$5.8$ event a day corresponding to a rate ${\Gamma\simeq 6.7
10^{-5}} s^{-1}$: the probability to find by chance one neutrino
event within a $1-2$ minute ${\Delta}t \simeq 10^2 s$ in that
interval is $P\simeq\Gamma \cdot{{\Delta}T}\simeq 6.7
\cdot10^{-3}$. For a Poisson distribution the probability to find
$n=1,2, 3, 4, 5$ events in a  narrow time window might reach
extremely small values:
$
 {{P}_{n}}\cong e^{-P}\cdot \frac{P^{n}}{n!}\simeq \frac{P^{n}}{n!}=( 6.7
\cdot 10^{-3}, 2.25 \cdot 10^{-5},  5 \cdot 10^{-8}, 8.39
10^{-11}, 1.1\cdot10^{-13}). $ Therefore just a few events
correlated to a Solar Flare will be a meaningful signal opening a
window to  novel Neutrino Astronomy.


  This paper is dedicated to the memory of the  flaring life of  Jonathan Evron,
   fallen  at brightest  age 20 on $2nd$ November 2005 for the Jewish State of Israel.


\end{document}